# An Accurate Process Induced Variability Aware Compact Model-based Circuit Performance Estimation for Design-Technology Co-optimization

Shubham Patil, Amita Rawat, and Udayan Ganguly *Senior Member, IEEE*

*Abstract* — In sub-10nm FinFETs, Line-edge-roughness (LER) and metal-gate granularity (MGG) are the two most dominant sources of variability and are mostly modeled semi-empirically. In this work, compact models of LER and MGG are used. We show an accurate process-induced variability (PIV) aware compact model-based circuit performance estimation for Design-Technology Co-optimization (DTCO). This work is carried out using an experimentally validated BSIM-CMG model on a 7nm Fin-FET node. First, we have shown performance benchmarking of LER and MGG models with the state-of-the-art and shown ~4x(~ 2.3x) accuracy improvement for NMOS(PMOS) in the estimation of device figure of merits(DFoMs). Second, RO and SRAM circuit's performance esti-mation is carried out for LER and MGG variability. Further, ~22% more optimistic estimate of $(\sigma/\mu)_{SHM}$ (Static Hold Margin) compared to the state-of-the-art model with $V_{DD}$ variation is shown. Finally, we demonstrate our improved DFoMs accuracy translated to more accurate circuits figure of merits (CFoMs) performance estimation. For worst-case SHM ($3(\sigma/\mu)_{SHM}$@$V_{DD}$=0.75 V) compared to state-of-the-art, dynamic(standby) power reduction by ~73%(~61%) is shown. Thus, our enhanced variability model accuracy enables more credible DTCO with significantly better performance estimates.

*Index Terms*— FinFET, Process induced variability (PIV), Design-Technology Co-optimization (DTCO), line-edge-roughness (LER), metal-gate-granularity (MGG), TCAD, BSIM-CMG, SPICE simulation, SRAM

## I. INTRODUCTION

AGGRESSIVE scaling improves performance but aggravates process induced variability. Statistical device-to-device variation caused by undesirable Process Induced Variability (PIV) sources produce circuit-level variations. To meet all specifications simultaneously for the distribution of devices, peak performance is compromised [1], [2].

In FinFET, LER and MGG have been shown to be the two most dominant sources of variability (Fig. 1) [4]. In most

This work is partially funded by Department of Science and Technol-ogy (DST) India, Indian Institute of Technology Bombay Nano-fabrication (IITBNF) facility IIT Bombay and Centre for Excellence (CEN) IIT Bom-bay.
Shubham Patil, Amita Rawat and Udayan Ganguly are with the Department of Electrical Engineering, Indian Institute of Technology Bombay, India (e-mail: udayan@ee.iitb.ac.in).

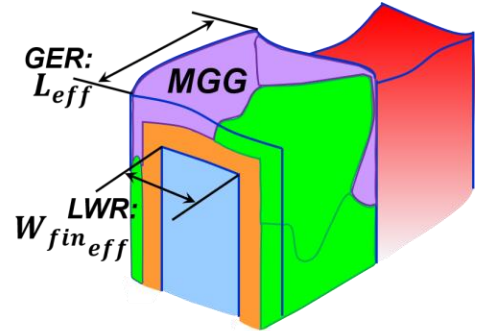

Fig. 1: Schematic of the FinFET with PIVs i.e., LER and MGG. Workfunctions of TiN used for $\sigma V_T$ calculation is shown in the Table I.

TABLE I: MGG in TiN [3]

| Orientation | WF | Probability |
|---|---|---|
| <100> | 4.6 eV | 60 % |
| <111> | 4.4 eV | 40% |

advanced nodes, Titanium Nitride (TiN) is the most commonly used gate metal in high-$k$ metal gate (HKMG) stack technology. The MGG variation in TiN is due to the presence of two different workfunction 4.4eV and 4.6eV with the probability of occurrence as 40% and 60% in <111> and <100> orientation respectively (Table I) [3]. The LER arises from fabrication induced fin shape variation. It is further categorized in two forms: 1) gate edge roughness (GER) and 2) fin edge roughness (FER). GER and FER accounts the variation in gate length and fin width respectively. The LER variability is characterized using input process dependent parameters such as auto-correlation length ($\Lambda$) and correlation coefficient ($\rho$) and root mean square (RMS) roughness ($\sigma$).

To introduce these LER variability information in circuit simulation, earlier, a semi-empirical model to capture the impact of LER on device performance has been reported in the literature [6]. However, it does not account for fin-edge correlation ($\rho$) which is significant in self-aligned quadruple patterning (SAQP). For MGG, the model in [7] depends on grain statistics without detailed positional dependence.

TABLE II: Benchmarking of our group LER and MGG models with state-of-the-art

| | Models | Fin statistics | | $V_T$ | Input Parameter |
|---|---|---|---|---|---|
| | | $\sigma_W$ | $\mu_W$ | | |
| LER | X.Jiang [6] | Semi-analytically calculated | | LUT (SPICE) | $\Lambda, \sigma$ |
| | Amita [8] | Compact form equation derived | | | $\Lambda, \sigma, \rho$ |

| | Models | Type | Size dependence/ Grain position |
|---|---|---|---|
| MGG | S. H. Rasouli [7] | Physical model | Yes/No |
| | H. Vardhan [9] | Analytical model | Yes/Yes |

TABLE III: Device dimension details [5]

| Type/Parameter | $L_G$ (nm) | $W_{fin}$ (nm) | $H_{fin}$ (nm) |
|---|---|---|---|
| NMOS | 12 | 5 | 60 |
| PMOS | 12 | 5 | 60 |

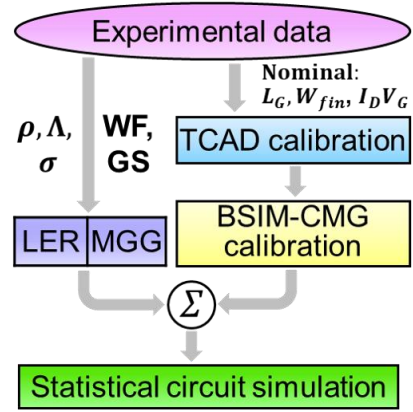

Fig. 2: Flow chart for model calibration, validation, and circuit simulation methodology adopted is shown.

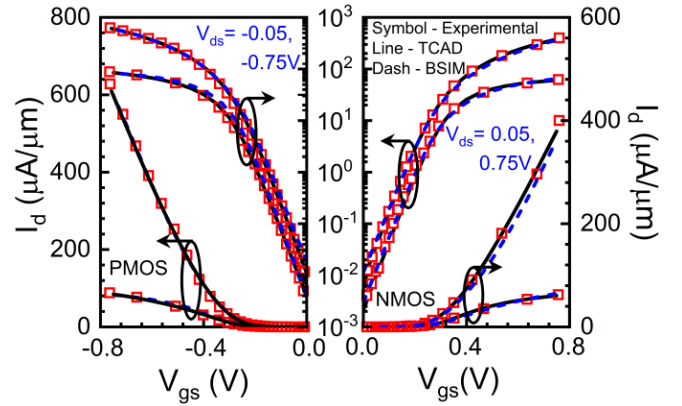

Fig. 3: Comparison of experimental data with the well calibrated TCAD deck and SPICE model of nominal device for PMOS and NMOS.

Recently, a compact model for LER [8] that is physics-based and accounting for all the critical variability parameters is proposed by our group. Further, a physics-based analytical model for MGG [9] considering the grains size and positional dependence work function is also proposed. Table II summarizes the model's comparison with state-of-the-art. A SPICE platform including [8], [9] was experimentally validated on 14nm node [10] in our earlier work.

In this work, we have extended the previously proposed PIV-aware SPICE framework for state-of-the-art 7nm CMOS technology FinFET to study the translation of device model accuracy to improvement in circuit performance estimation. At first, to ensure the accuracy of the results, we have calibrated the TCAD deck against the experimental data presented in [11] by GF for their 7nm FinFET technology. The calibrated TCAD deck is used to perform an elaborate calibration of the BSIM-CMG model [12]. Second, we implement our model and compare it against the state-of-the-art semi-analytical model [6]. Finally, we demonstrate our improved accuracy translated to significantly different and hence more accurate performance estimation in the typical circuits - ring oscillator (RO) and SRAM using our PIV-aware SPICE framework.

This paper is organized as follows. The device details and simulation parameters are discussed in Section II. The TCAD and SPICE calibration is shown in section III-A and III-B followed by benchmarking of PIV-aware simulation framework with TCAD in III-C. Impact of PIVs on typical circuits performance - RO and SRAM are discussed in section IV-A and the comparison of accuracy in performance estimation with state-of-the-art is discussed in section IV-B. The impact of improved accuracy in device variability estimation on circuit performance is discussed in section IV-C followed by conclusion in section V.

## II. DEVICE DETAILS AND SIMULATION PARAMETERS

Fig. 1 shows the schematic of the FinFET with PIVs. The device dimensions [5] used in this work are mentioned in Table III. An industry standard BSIM-CMG model is adopted for SPICE simulation. The BSIM-CMG is calibrated with the experimental data and elaborate calibration is done using statistical simulation data from an experimentally calibrated TCAD deck. $\sigma V_T$ values for process variation induced average grain size of 4 nm and $\sigma_{LER}$ of 2 nm are considered in SPICE simulation unless specified. Typical circuits - RO and SRAM circuit performance analysis is carried out across variation of different LER parameters: $\rho$, $\Lambda$ and $\sigma$. Similarly, for MGG, $\sigma V_T$ values corresponding to different TiN grain sizes are calculated using the analytical model presented in [9] for the device dimensions mentioned in Table III. Further, those $\sigma V_T$ values are used in the SPICE simulation.

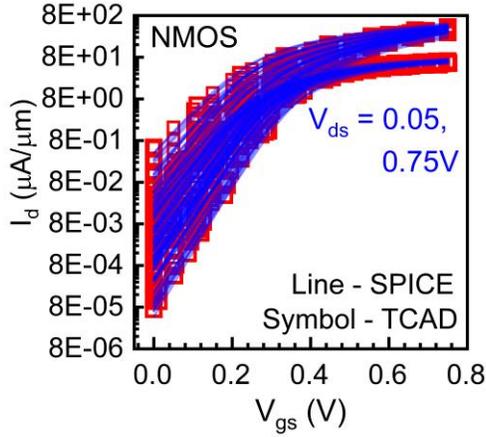

Fig. 4: Comparison of $I_d$-$V_{gs}$ data of statistical TCAD simulation against SPICE simulation. Good consistency between the two conforms the PIVs aware framework capability to capture the LER and MGG variations well.

TABLE IV: TiN grain size dependent $\sigma V_T$ calculation [9]

| Grain size (nm) | 3 | 3.5 | 4 | 4.5 | 5 |
|---|---|---|---|---|---|
| $\sigma V_T$ (mV) | 25 | 27.5 | 33 | 39 | 47 |

## III. CALIBRATION AND BENCHMARKING OF THE DEVICE VARIABILITY MODEL

In this section, the TCAD and SPICE calibration is discussed, followed by the validation of the SPICE framework against calibrated TCAD deck using statistical simulation. The DTCO flow is summarized in Fig. 2.

### A. TCAD calibration

In TCAD Sentaurus [13] test bench, the following device physics models are used. To capture the current transport, drift-diffusion model is used with density gradient model to include the quantum correction (QC) [14]. The mobility models included are (1) IALMOB model [15] to capture mobility degradation due to doping and Enormal component. While thin layer model [16] is used in conjunction with IALMOB model to capture mobility degradation due to geometric quantization, (2) extended Canali model [17] to capture high field saturation and (3) BALMob model [18] to account for ballistic transport. SRH recombination [19], trap assisted tunneling (TAT) [20] and band to band tunneling [21] models are used to capture the recombination and generation current. Using the above models, the TCAD deck is calibrated with the experimental data and the excellent matching is shown for the nominal device in Fig. 3 for both NMOS and PMOS. Further, the calibrated 2-D simulation deck is used to generate the gate length ($L_G$) and fin width ($W_{fin}$) splits for the elaborate calibration of BSIM-CMG model across the geometric variation. The BSIM-CMG calibration and validation is discussed in the following section.

### B. SPICE calibration

We have calibrated the BSIM-CMG with the experimental data of the nominal device for NMOS and PMOS. The calibration is extended for the range of $L_G$ and $W_{fin}$ using data generated from calibrated TCAD deck. The model parameters used for calibration are discussed below. First, device physical parameters are fixed to define device geometry. Further, to capture the threshold voltage, the workfunction of the device is tuned using PHIG. Second, U0 and UA parameters for low field mobility and phonon/surface roughness scattering are used to capture the mobility degradation. While the velocity saturation effects are captured using VSAT and VSAT1 parameters. Third, the short channel effects are captured using DVT0 and DVT1, and PCLM is used to capture the channel length modulation. In the sub-10nm regime, small geometry-induced effect, i.e., quantum confinement, became severe. QMFACTOR is enabled in the model for quantum correction. DVT1SS, U0 and VSAT are used for subthreshold swing(SS), mobility and saturation velocity correction respectively for calibration across different device geometries. The BSIM-CMG model calibration for the nominal device data is shown in Fig. 3. In the next section, the benchmarking of the SPICE framework with calibrated TCAD is discussed based on the BSIM-CMG calibration.

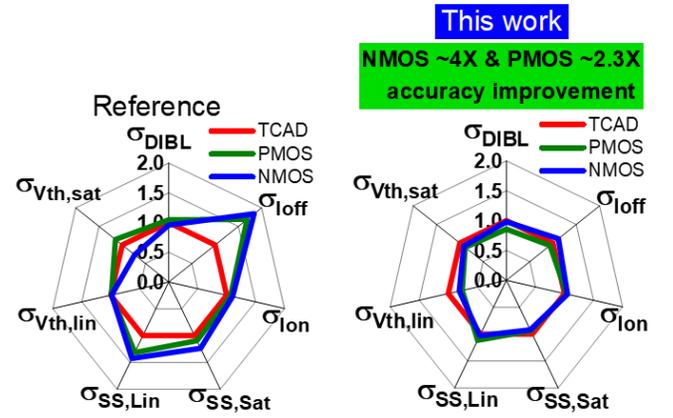

Fig. 5: Benchmarking of PIV-aware framework accuracy in estimation of PIV impact on DFoMs with respect to TCAD is shown. All $\sigma$'s are rationalized with the respective $\sigma_{TCAD}$. This work shows ~ 74% (~ 54%) $\rho_{accuracy}$ improvement for NMOS and PMOS as compared to referenced model [6].

### C. Benchmarking of the device variability model

Using the calibrated BSIM-CMG model, statistical simulations are performed on the developed PIV-aware framework in the SPICE. SPICE simulation shows well consistency with the statistical data from calibrated TCAD as shown in Fig. 4. Further, to compare the accuracy in estimation of device figure of merits (DFoMs): $SS_{lin}$, $SS_{sat}$, $V_{th,lin}$, $V_{th,sat}$, $I_{on}$, $I_{off}$ and DIBL due to PIVs, the standard deviation ($\sigma$) are calculated from statistical data of TCAD and SPICE. Comparison of calculated $\sigma$ of DFoMs for both NMOS and PMOS with respect to TCAD is shown using spider plot in Fig.5(a) and 5(b). All $\sigma$'s are normalized with the respective $\sigma_{TCAD}$.

This work shows ~4x(~2.3x) $\sigma_{error}$ reduction for NMOS(PMOS) as compared to the referenced model [6]. This developed framework is used for performance estimation of RO and SRAM circuits, discussed in the next section.

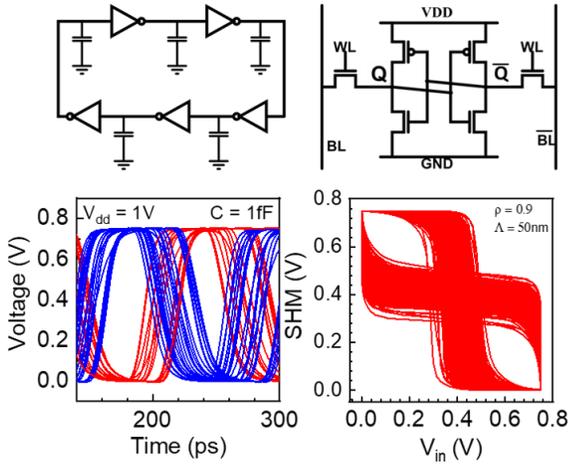

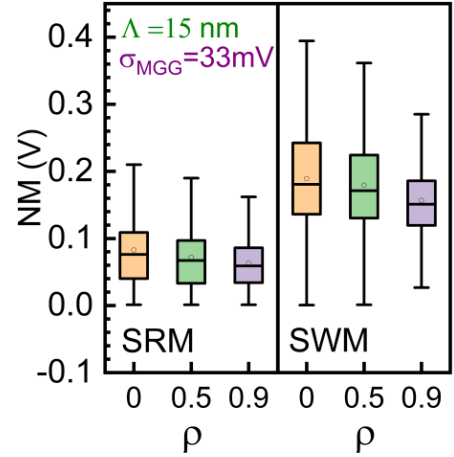

Fig. 6: Circuits (a) 5 stage RO with capacitor of 1fF and (b) 6T SRAM are used for demonstration. The outputs are shown for statistical simulation using the developed PIV aware framework.

Fig. 8: SRAM Static read margin (SRM) and Static write margin (SWM) comparison across different $\rho$. With the increase in $\rho$, SRM (SWM) degrades by ~23%(~17%) for $\Lambda$ = 15 nm. While the increase in $\rho$, shows the decrease in variability.

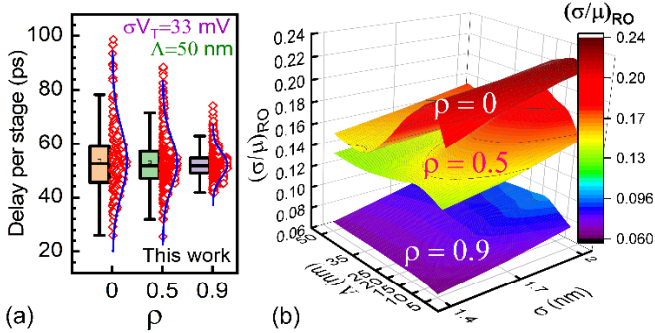

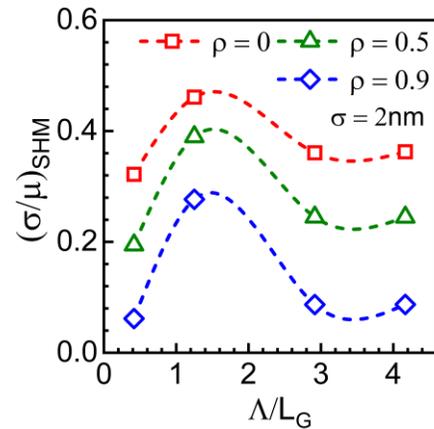

Fig. 7: (a) RO delay comparison (at fixed $\Lambda$=50 nm) across different $\rho$. With the increase in $\rho$ ($\rho$=0.9 as compared to $\rho$=0), delay shows a negligible (~2%) decrease. While $(\sigma/\mu)_{RO}$ decreases by (~61%). (b) $(\sigma/\mu)_{RO}$ comparison across different LER parameters: $\rho$, $\Lambda$ and $\sigma$ variation for fixed $\sigma V_T$ = 33 mV. With an increase in $\rho$, $(\sigma/\mu)_{RO}$ decreases at fixed $\Lambda$ and $\sigma$. While decrease in $\sigma$ shows monotonic decrease in $(\sigma/\mu)_{RO}$.

Fig. 9: Dependence of $(\sigma/\mu)_{SHM}$ on $\Lambda/L_G$ is shown for different $\rho$. $(\sigma/\mu)_{SHM}$ increases with decrease in $\Lambda$ and peaks as $\Lambda/L_G$ tends to ~1 and reduces on further reduction of $\Lambda/L_G$. Also, $(\sigma/\mu)_{SHM}$ decreases with $\rho$, due to increase correlation between edges, leading to lower variability (results are qualitatively consistent with referenced model [6]).

## IV. RESULTS AND DISCUSSION

### A. RO and SRAM Circuit performance analysis

To study the impact of variability and fabrication aspects, (a) RO and (b) 6T SRAM cell are used as shown in Fig. 6, along with the RO characteristics and static hold margin (SHM) of SRAM. RO and 6T SRAM cells performance are analyzed across different lithography techniques such as fin edge correlation coefficient $\rho$ = 0 for EUV, $\rho$ = 0.5, 0.9 for SADP [10] and across different $\Lambda$ = 5, 10, 15, 20, 25, 35, 50 nm and $\sigma$ = 1.4, 1.7 and 2 nm. Similarly, for MGG, different TiN grain sizes of 3, 3.5, 4, 4.5 and 5 nm are considered for $\sigma V_T$ calculation (Table IV) [9].

**RO**: Fig. 7(a) shows a negligible (~2%) decrease in RO delay for $\rho$ = 0.9 compared to $\rho$ = 0. But there is a significant decrease in variability for higher $\rho$. Also, $(\sigma/\mu)_{RO}$ analysis is carried out across $\rho$, $\Lambda$ and $\sigma$ for constant grain size of 4nm as shown in Fig. 7(b). It is observed that $(\sigma/\mu)_{RO}$ decreases with increase in $\rho$ and decrease in $\sigma$. Also, the $(\sigma/\mu)_{RO}$ peaks when $\Lambda/L_G$ tends to ~1.

**SRAM**: Fig. 8 shows the SRAM (a) the static read margin (SRM) and (b) static write margin (SWM) degrade by ~23% (~17%) at $\Lambda$ = 15 nm for $\rho$ = 0.9 as compared to $\rho$ = 0. Fig.9 shows the dependence of $(\sigma/\mu)_{SHM}$ on $\Lambda/L_G$ for different $\rho$. $(\sigma/\mu)_{SHM}$ increases with decrease in $\Lambda$ and peaks as $\Lambda/L_G$ tends to ~1 and reduces on further reduction of $\Lambda/L_G$, the results are qualitatively consistent with referenced model [6]. The SHM is further analyzed across different LER parameters as shown in Fig. 10. It has been observed that $(\sigma/\mu)_{SHM}$ decreases for higher $\rho$, due to an increase in correlation between edges (reduced LER), leading to lower variability. It also reduces with the decrease in $\sigma$.

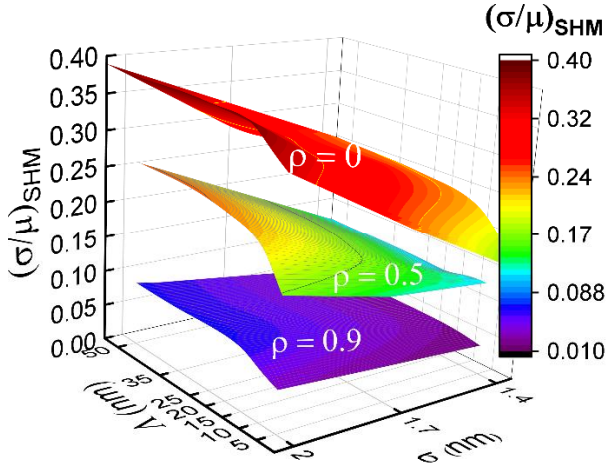
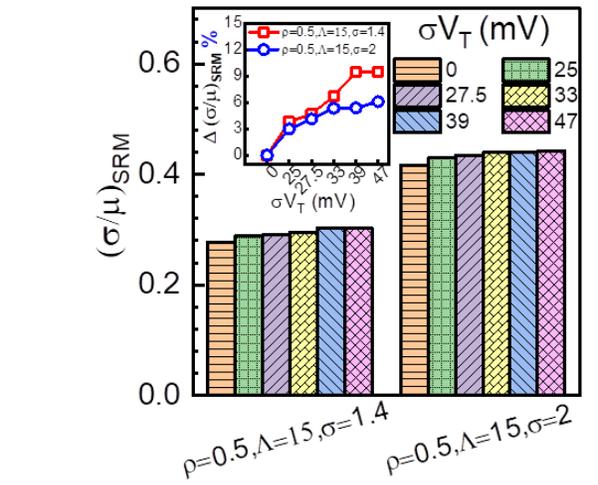

Fig. 10: Dependence of $(\sigma/\mu)_{SHM}$ on $\Lambda$ and $\sigma$ for different $\rho$. Variation is highest as $\Lambda/L_G \sim 1$ while it decreases with increase in $\rho$. It helps in establishing the design technology co-optimization (DTCO) strategy to choose suitable $\Lambda$ and use lithography techniques with higher $\rho$ and lower $\sigma$ to improve SRAM performance.

Fig. 11: Dependence of $(\sigma/\mu)_{SRM}$ on $\sigma V_T$ for fixed $\rho$ and $\Lambda$ and for $\sigma = 1.4$ and 2 nm is shown. Increase in $\sigma V_T$ leads to increase in $(\sigma/\mu)_{SRM}$ for both cases as shown in inset plot. The contribution due to MGG variation in $(\sigma/\mu)_{SRM}$ is not significant (~10% as compared to $\sigma V_T = 0$). Hence, LER is acting as a dominant source of variability.

The RO and SRAM analysis shows the $(\sigma/\mu)_{RO}$ and $(\sigma/\mu)_{SHM}$ peaks as $\Lambda/L_G$ tends to 1. This is because the initial increment in $\Lambda$ makes the fin edges more regular, which increases the $\sigma$ and $\mu$, however, the increment rates are different for $\sigma$ (faster) and $\mu$ (lower). Moreover, as $\Lambda/L_G > 1$, the sigma saturates while the $\mu$ saturates as $\Lambda/L_G \sim 3$ [8].

To study the impact of MGG variability, SRM is analyzed for $\sigma V_T$ variation (Table IV) across different LER parameters as shown in Fig. 11. The inset shows with the increase in $\sigma V_T$, $\Delta(\sigma/\mu)_{SRM}$ given in Equation % (1) increase. It is noteworthy to mention that the contribution due to MGG variation in $(\sigma/\mu)_{SRM}$ is not significant (~10% as compared to $\sigma V_T = 0$). Hence, LER is acting as a dominant source of variability. Thus, the design technology co-optimization (DTCO) strategy by using lithography techniques with higher $\rho$ and $\Lambda$, and lower $\sigma$ to improve RO and SRAM performance is quantitatively established. In the next section, the comparison of accuracy difference in performance estimation with state-of-the-art is discussed.

$$\Delta\left(\frac{\sigma}{\mu}\right)_{SRM}\% = \frac{\left(\frac{\sigma}{\mu}\right)_{SRM}|_{\sigma V_T} - \left(\frac{\sigma}{\mu}\right)_{SRM}|_{\sigma V_T=0}}{\left(\frac{\sigma}{\mu}\right)_{SRM}|_{\sigma V_T=0}} \quad (1)$$

### B. Circuit performance estimation accuracy comparison

To check the accuracy in performance estimation as compared to the referenced model [6], SHM is analyzed for $V_{DD}$ variation as shown in Fig. 12(a). The $(\sigma/\mu)_{SHM}$ for both cases is shown in Fig. 12(b) and 22% improvement in $\Delta(\sigma/\mu)_{SHM}$ % (2) estimation is observed using our setup as shown in Fig. 12(c). Further, we quantify the operating conditions predicted by our model for the same performance as the referenced model [6]. We compared the worst case SHM, i.e., $3\sigma_{SHM}$ that we got from our model and the referenced model [6] as shown in Fig. 13. It shows a significant reduction in operating $V_{DD}$ (-180mV) is possible for the worst-case SHM window (for same $3\sigma_{SHM\ [6]}$@$V_{DD}$=0.75V).

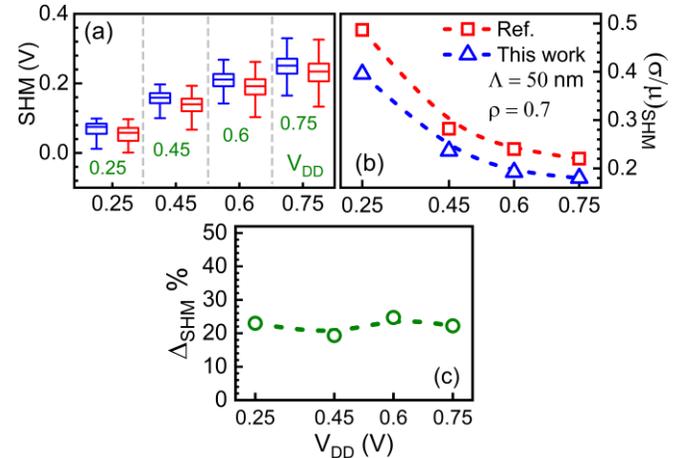

Fig. 12: (a) SHM comparison is shown with referenced model [6] across $V_{DD}$ variation. (b) Shows $(\sigma/\mu)_{SHM}$ comparison and ~22% more optimistic estimate of $(\sigma/\mu)_{SHM}$ is observed, as shown in (c).

$$\Delta\left(\frac{\sigma}{\mu}\right)_{SHM}\% = \frac{\left(\frac{\sigma}{\mu}\right)_{SHM[Ref]} - \left(\frac{\sigma}{\mu}\right)_{SHM[This\ work]}}{\left(\frac{\sigma}{\mu}\right)_{SHM[This\ work]}} \quad (2)$$

### C. Model accuracy translation to circuit performance

Fig. 14 shows the relative difference benchmarking in the DFoM and circuit figure of merits (CFoM): $(\sigma/\mu)_{SHM}$, standby and dynamic power benchmarked to the referenced model [6]. More accurate DFoMs estimation leads to the significant difference in CFoMs performance. Our model predicts a

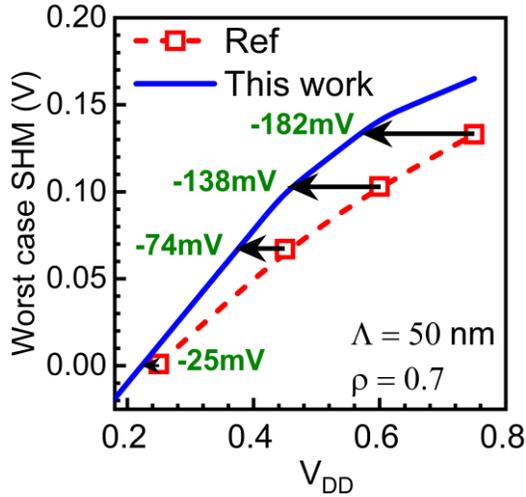

Fig. 13: Worst case SHM ($3\sigma_{SHM}$) comparison of this work with the referenced model [6] is shown for $V_{DD}$ variation. It is observed that this model enables significant $V_{DD}$ reduction for same performance. It can lead to a significant reduction in dynamic power.

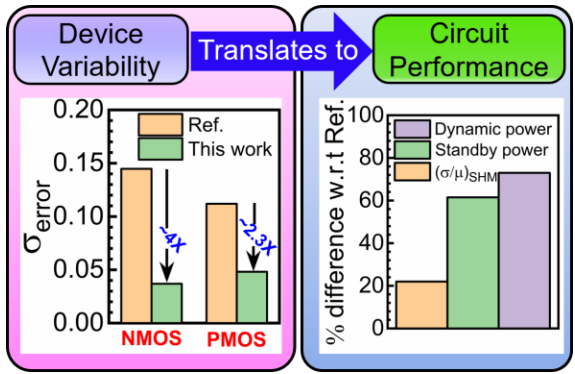

Fig. 14: Relative difference in comparison to referenced model [6] prediction is shown for DFoMs accuracy estimation translating to more optimistic CFoMs performance estimation, enabling the scope for circuit designers to further optimize performance.

~22% more optimistic estimate of SHM compared to the refrenced model [6], which enables more aggressive circuit design. Further, it allows the SRAM to operate at lower $V_{DD}$ (-180mV) for $3(\sigma/\mu)_{SHM\ [6]}$@$V_{DD}$=0.75 V, leading to dynamic power reduction by ~73%. Also, ~61% reduction in average standby power is observed.

## V. CONCLUSION

To summarize, we demonstrated the relative improvement in DFoMs estimation using our PIV aware SPICE simulation framework using in-house LER and MGG models c.f state-of-the-art [6]. The DFoMs accuracy enhancement leads to significantly more accurate circuit performance estimation in RO and SRAM that improves CFoMs estimates. Further, the impact of variability is studied across different technological parameters (LER: $\rho$, $\sigma$, $\Lambda$ and MGG: Grain size). Our platform enables accurate DTCO for optimized circuit performance.

## ACKNOWLEDGMENT

This work is partially funded by Department of Science and Technology (DST) India, Indian Institute of Technology Bombay Nano-fabrication (IITBNF) facility IIT Bombay and Centre for Excellence (CEN) IIT Bombay.